# Freezing and melting of vortex ice

J. Trastoy[1], M. Malnou[2], C. Ulysse[3], R. Bernard[1], N. Bergeal[2], G. Faini[3], J. Lesueur[2], J. Briatico[1] and Javier E. Villegas[1]

[1]Unité Mixte de Physique CNRS/Thales, 1 ave. A. Fresnel, 91767 Palaiseau, and Université Paris Sud 11, 91405 Orsay, France

[2]LPEM, ESPCI-CNRS-UPMC, 10 rue Vauquelin 75231 Paris, France.

[3]CNRS, Phynano Team, Laboratoire de Photonique et de Nanostructures, route de Nozay, 91460 Marcoussis, France.

We report on the realization of artificial ice using superconducting vortices in geometrically frustrated pinning arrays. This vortex ice shows two unique properties among artificial ice systems. The first comes from the possibility to switch the array geometric frustration on/off through temperature variations, which allows "freezing" and "melting" the vortex ice. The second is that the depinning and dynamics of the frozen vortex ice are insensitive to annealing, which implies that the ordered ground state is spontaneously approached. The major role of thermal fluctuations and the strong vortex-vortex interactions are at the origin of this unusual behavior.

Vortices in type-II superconductors constitute a model to investigate the physics of "interacting particles" on energy landscapes. The possibility to design the vortex energy landscape via artificial arrays of pinning centers has allowed the study of a flood of problems (commensurability [1], jamming [2], rectification [3], avalanches [4], etc) which concern a heterogeneous variety of systems (colloids [5], cold atoms in optical traps [6], swimming bacteria [7], proteins in motion [8], electrons in Wigner crystals [9], etc). One interesting problem is geometric frustration, which appears when the constraints imposed by the energy landscape do not allow all particles to simultaneously minimize their pairwise interaction energy. This scenario is found in various natural systems, the archetypal being water ice [10] and pyrochlore crystals [11] (in which an analogous "spin ice" forms). Artificial ice systems attract increasing attention [12,13,14,15] as they allow the study of frustration under controlled conditions. Numerical simulations [16] have proven superconducting vortices a promising system to realize artificial ice, and experimental realizations are underway [17,18].

So far, most of the artificial ice realizations have been done using 2D arrays of nanomagnets that present bi-stable single-domain magnetization, in which the pairwise dipolar interactions are minimized when magnetizations are head-to-tail [12,15,19,20,21]. Geometric frustration allows minimizing only a fraction of those interactions, and the lowest-energy state is achieved when magnetizations obey the so-called ice rule [10] at the array vertices [12]. However, the ordered ground state has proven difficult to observe experimentally in this type of systems. This is because of the relative weakness of the magnetic dipolar interactions, and because the energy barrier between the two possible magnetization states of each nanomagnet is typically much higher than the thermal energy during experiments [20]. This situation usually leads to frozen static systems [20, 22]. Thus, except for systems annealed during fabrication [20], extended nanomaget arrays show a significant amount of disorder: while a statistical preference for ice-rule-obeying vertices is observed, many other vertices have high-energy configurations forbidden by the ice rule [12]. Only small-sized systems (two to three vertices) [21] show thermal behavior and spontaneously approach the ground state. However, such small-sized systems lack long-range interactions, whose role in ice dynamics and ordering is an open issue [23].

In this Letter, we experimentally realize vortex ice in high-temperature superconducting thin films with artificial pinning arrays. A rare characteristic of this extended ice system is the role played by temperature. In particular, we show that varying the temperature allows one to switch "on" and "off" the geometric frustration of the energy landscape. This enables us to controllably "freeze" and "melt" vortex ice, a unique possibility among artificial ice systems. Furthermore, we find that the depinning and dynamics of vortex ice is insensitive to annealing, which suggests that the ice-rule-obeying ground state (or a very close one) is approached spontaneously. The keys to this unusual behavior are the use of a high-temperature superconductor −in which vortex thermal fluctuations are large− and of arrays in which the distance between pining sites is much shorter than the superconducting penetration length, which leads vortices to strongly interact.

The pinning arrays are defined in 50 nm thick $YBa_2Cu_3O_{7-\delta}$ films grown on $SrTiO_3$ via $O^+$ ions irradiation (energy 110 keV, fluence $5·10^{13}$ ions·cm$^{-2}$) through nano-perforated masks. These are defined by e-beam lithography in a thick resist (PMMA) spin-coated on the YBCO surface (see details elsewhere [24]). Fig. 1(a) shows one of the masks' micrograph. The hole arrays are defined with a resolution better than ~10 nm, which is crucial since minute variations of the array geometry on this scale dramatically change the observed behavior (see below). Masked ion irradiation induces point defects in the YBCO, which locally depress its critical temperature $T_C$ [25]. The local $T_C$ distribution can be estimated from Monte Carlo simulations of the induced defect concentration, considering the Abrikosov-Gorkov depairing law [25]. Examples of the calculated local $T_C$ maps are show in Figs. 1(b)-(d). $T_C$ is markedly depressed in the circular areas exposed to the ion beam. These areas behave as strong pinning sites for the vortices [24]. The samples' superconducting transition, as determined from R(T), is characterized by an onset around ~86 K and a zero-resistance $T_{C0}$ ranging 63 K-75 K depending on the mask (see Supplementary Information).

The geometry of the arrays used in these experiments is shown in Fig. 1. The holes' diameter is ∅~70 nm. We studied arrays with different parameters $L_1$ and $L_2$ [see definition in Fig. 1(b)]. Here we discuss a series with fixed $L_2$=120 nm and 60 nm<$L_1$<150 nm. Each array is characterized by $B_\phi = \phi_0/(L_1 + 2L_2 \cos 45)$, the field at which there is one flux quantum per array unit cell (shown in Fig. 1(a), dashed square). The array geometric frustration is evident in the situation depicted in Fig. 1(a), in which there are two vortices (blue circles) per unit cell: if one tries to place every vortex in a pinning site, there is no possible distribution in which the distance between the strongly repulsive vortices is constant. It is therefore not possible to minimize the pairwise interaction energy for every vortex. The best vortices can do to globally minimize the interaction energy is to order obeying the ice-rule as in Fig. 1(a): two vortices occupy pining sites closer to the vertex (dashed circle), and two other occupy outer positions, alternating every array vertex. This corresponds to the square-ice ground state [16]. The question is whether this is actually achieved, or whether vortices prefer to keep their natural (triangular) ordering at the expense of not being located in the pinning sites.

Figs. 2(a)-(d) show the magneto-resistance of the series of samples. The magnetic field $B$ is expressed in terms of the number of vortices per unit cell $n = B/B_\phi$. A SEM image of each PMMA mask is displayed, and $L_1$-$L_2$ (in nm) indicated. For each curve, the reduced temperature $t=T/T_{C0}$ and the injected current density $J$ in A/m$^2$ are indicated.

First we compare the highest temperature curves ($t$~1, red lines) in Figs. 2(a)-(d). These measurements are done in the (low-current) Ohmic regime above the irreversibility line [26]. Note that for the array in Fig. 2(a) circular holes have merged by pairs into ovals because $L_1$=60 nm<∅. Thus, this array is not geometrically frustrated, but just a square array with oval pinning sites. It will be used for comparison. The magneto-resistance curve of this sample shows deep minima for even values of |n| (the deepest for |n|=2), and much shallower for odd values. Notably, the 90-120 array behaves similarly [Fig. 2(b)], despite presenting geometrical frustration. The two other geometrically frustrated arrays ($L_2 \geq L_1$) have strikingly different behaviors. For the 120-120 one [Fig. 2(c)], the deepest minima are for |n|=4, whereas the other minima are much shallower (|n|=2) or barely visible (|n|=1, 3). Finally, for the 150-120 array [Fig. 2 (d)] the most pronounced minima correspond to |n|=1, while higher order minima gradually become shallower, and those corresponding to |n|=4 are barely noticeable.

When *t* is decreased, a gradual smoothing of the curves is observed in all cases. However, a closer look reveals very different temperature effects on the minima's relative depth. For the arrays in Figs. 2(a) and 2(c), no temperature evolution is observed, so that respectively the minima at |n|=2 and |n|=4 remain the most pronounced ones at any temperature. Contrary to this −and surprisingly− in Figs. 2(b) and 2(d) the minima for |n|<4 are smoothed away much more markedly than the minima at |n|=4. In this manner, at the lowest temperature (blue curves) the minima for |n|=4 become the most pronounced ones, and the others are barely noticeable (|n|=2) or completely washed out (|n|=1, 3). Thus, despite the great differences at high *t*, Figs. 2(b)-(d) become very similar at low *t*. Conversely, Figs. 2(a) and 2(b) become very different upon cooling despite their similarity at high *t*. Such a dramatic modification of the field-matching effects upon cooling is very unusual, and suggests that the vortex energy landscape geometry strongly depends on temperature.

The lowest *t* curves in Fig. 2 were measured below the irreversibility line, where the critical depinning current $J_C$ is well defined [26]. Thus, we can use $J_C$ *vs.* the applied magnetic field B to investigate the static ordering of vortices in this regime. $J_C(B)$ was obtained from sets of isothermal E-J characteristics measured at t~0.8 in a series of increasing magnetic fields. A criterion $E_C=2.5·10^{-2}$ V/m was used to determine $J_C$. The matching effects in $J_C(B)$ [Fig. 3(a)] parallel those observed in the low-*t* magneto-resistance curves. For the three samples, the most pronounced peak corresponds to n=4, and a weaker one is observed for n=2. Note however that the latter is well-defined for the 90-120 array, barely noticeable for 120-120, and unnoticeable for the 150-120 array [same trend as in the low-*t* curves in Figs. 2(b)-(d)]. In order to quantitatively describe this behavior, we show in Fig. 3(b) the relative $J_C$ enhancement at the matching condition. This is defined as the $J_C$ peak height $\Delta J_C$, [defined in Fig. 3(a)] relative to the background (variation of $J_C$ in the experimental window), i.e. $\Delta J_C/(J_C(0) - J_C(7B_\Phi))$. Note that, for both peaks n=2 and n=4, the $J_C$ enhancement steadily decreases as the array parameter $L_1$ increases from 90 to 150 nm.

Let us now explain the origin of the observed behaviors. For the sample in Fig. 2(a), understanding the series of minima is straightforward: the ovals form a square array of pinning sites, for which field-matching effects are well known. The minima at |n|=2 correspond to *one* flux quantum per pinning site [see sketch], those at |n|=4 correspond to *two*, etc. The shallow dips at |n|=1 and |n|=3 correspond to the so-called fractional matching, in which there is a non-integer number of flux quanta per pinning site. For half a flux quantum per pinning site (|n|=1), for example, vortices leave half of the sites empty and form an ordered checkerboard pattern [27]. The shallowness of the minima at fractional matching, also observed earlier [28,29,30], can be understood considering that vortices in motion jump from a pinning site to the closest one [31] pushed by the current (Lorentz force) and/or assisted by thermal activation. For fractional matching, vortices find the closest pinning sites empty, whereas for integer matching, vortices find all the closest pinning sites occupied by their (strongly repulsive) nearest neighbors. In the latter situation, vortex depinning requires the synchronized depinning of the nearest neighbor. This explains why fractional matching is less stable, more prone to disorder [27] and presents a weaker pinning enhancement than integer matching [28,29,30].

We turn now to Fig. 2(b). The high-*t* curve's resemblance with that in Fig. 2(a) suggests that the vortex energy landscape is similar in both samples. This can be understood if one considers vortex thermal activation, which leads to a smoothing of the artificial pinning landscape [26]. Fig. 1(b) shows that the $T_C$ depression between pinning sites is stronger along $L_1$ (shortest) than along $L_2$ (longest). Thus, one expects that at high enough *t* the thermal smoothing of the energy landscape makes the barrier across $L_1$ negligible as compared to that across $L_2$, so that the sites separated by $L_1$ virtually "merge" [ovals sketched in Fig. 2(b)] while vortices still see a barrier across $L_2$. In consequence, at high *t*, the vortex energy landscape is virtually similar to the 60-120 sample [Fig. 2(a)], and the magneto-resistance can be analogously explained in terms of the square-array behavior.

The high-*t* curve in Fig. 2(d) can be understood via the same arguments. In this case $L_2<L_1$, and therefore one expects the four holes circled in the sketch for n=1 to "merge" at high *t*, virtually yielding a square array of large "composite" pinning sites. This explains that the most pronounced minima correspond to |n|=1 (one flux quantum per "composite" site), while higher order minima gradually become shallower, as expected for square arrays [29]. Note however that the situation is different for the sample in Fig. 2(c), for which $L_1=L_2$ and all energy barriers between pinning sites are the same [see Fig. 1(c)]. Thus, temperature effects analogous to those in Figs. 2(b) and 2(d) are not anticipated, and indeed not observed in Fig. 2(c).

When thermal activation diminishes, the virtually merged pinning sites split up, yielding similar energy landscapes for all 90-120, 120-120 and 150-120 samples. This is evidenced by the similar low-*t* curves in Figs. 2(b)-(d) and $J_C$(B) in Fig. 3(a). All the curves show the same hallmark: matching effects are observed only for |n|=2 and |n|=4, the latter being the most pronounced. For |n|=4 there are as many vortices as pinning sites [sketched in Fig. 2(d)]. For |n|=2, half of the pinning sites are empty, and vortex ice forms. The question is whether the square-ice ground state sketched in Fig. 1(a) is actually achieved, or if a sizable amount of disorder exists (vortices not obeying the ice-rule) which may account for the weakness of the matching effects at |n|=2 and their varying strength for the different arrays [Fig. 3(b)].

In order to look for disorder effects, we "annealed" the vortex lattice to see whether this would enhance matching effects around |n|=2. The idea was to repeatedly "shake" vortices back and forth in order to reposition those not obeying the ice-rule, so as to approach the ground state. We used different "annealing" protocols (see details in the Supplementary Information), including the injection of rotating currents earlier proposed [16], radio-frequency excitation, etc. However, we observed no difference between the V(I) nor R(B) after and before annealing. From this, we ruled out the presence of a significant amount of disorder in the vortex ice. We concluded that the |n|=2 matching effects are weaker than for |n|=4 simply because they are a fractional matching state (half of the pinning sites empty), and the same arguments used above for square arrays apply.

We discuss below the gradual weakening of the low-*t* matching effects as $L_1$ is increased from 90 to 150 nm, a trend observed in Figs. 2 and 3(b). For the vortex lattice to match the pinning array and produce a $J_C$ enhancement, the lattice needs to deform from its natural geometry (triangular) into the array's one. This happens if the elastic energy increase due to the deformation $\Delta E_{el}$ is outbalanced by the pinning energy $\Delta E_{pin}$ [32]. Therefore, for the matching to occur, $\Delta E_{el} < \Delta E_{pin}$ must be fulfilled, and the stronger the inequality, the more stable the vortex configuration, and the stronger the $J_C$ enhancement [33]. $\Delta E_{el} = E_{int}^{tr} - E_{int}^{m}$, where the two subtracted terms are the interaction energy between vortices for the triangular (tr) and matching (m) geometries. In the high-$\kappa$ limit, $E_{int} = \phi_0^2/4\pi\lambda^2\mu_o \sum_j K_0(r_{0j}/\lambda)$ [33,34], with $\lambda$ the temperature dependent penetration depth, $\mu_o$ the magnetic permeability of the vacuum, $r_{0j}$ the position of the $j$-th vortex, and $K_0$ the zeroth-order modified Bessel function of the second kind. We have used $\lambda = (\lambda_0^2/d) \cdot (1 - t^4)^{-1/2}$, with $\lambda_0$=150 nm and $d$=50 nm the thickness of the film [33]. $\Delta E_{pin} = (v_p^m - v_p^{tr}) \cdot \varepsilon_p$, with $v_p$ being the fraction of pinned vortices for each geometry, and $\varepsilon_p$ the pinning energy per site. As detailed in the Supplementary Information, $v_p^{tr} \in [0.11,0.17]$, depending on L$_1$. By definition, $v_p^m$=1. $\Delta E_{pin}$ only changes moderately when L$_1$ is changed [see Fig. 3(c)]. However, the elastic energy strongly increases with increasing L$_1$ (square symbols). This means that the balance $\Delta E_{el} < \Delta E_{pin}$ becomes less favorable as L$_1$ increases, yielding weaker matching effects, as observed [Fig. 3(b)].

In summary, we have studied vortex ordering in geometrically frustrated "square-ice" pinning arrays. At low temperatures, all the arrays behave similarly. In this regime, the field-matching effects observed in the magneto-transport −and their intensity as a function of the array parameters− can be consistently understood considering that all the vortices localize in array pinning sites and are confronted to the geometric frustration, thereby forming a square vortex ice for the appropriate vortex density (n=2). The fact that the magneto-transport properties are insensitive to annealing implies that the ground state is spontaneously approached. For the frustrated arrays in which two different distances/energy barriers between nearest pining sites exist, increasing the temperature dramatically changes the energy landscape geometry, and washes out geometric frustration at high temperature. This way, the vortex ice, stabilized at low temperatures, "melts" at higher ones and becomes a periodic (square) vortex lattice. Because of the major role played by temperature and the possibility to finely tailor the energy landscape, we believe that the system realized here holds much potential for further studies on the dynamics and thermal properties of artificial ice. Issues such as relaxation, hysteretic behavior and size effects are naturally arising questions, for which the combination of transport and imaging techniques [35] may be key.

Work supported by French ANR « SUPERHYBRIDS-II » and « MASTHER » grants, and the "Ville de Paris Emergence Programme". J. T. acknowledges the support of Fundación Barrié (Galicia, Spain).

See Supplemental Material at [URL will be inserted by publisher] for details when pointed in the main text.

**Fig. 1 (a)** Scanning electron microscopy image of a 150-120 mask used for $O^+$ ion irradiation. The (yellow) dashed circle indicates an array vertex and the dashed square the array unit cell. Blue circles represent vortices. **(b)**-**(d)** Simulated local critical temperature map (scale in K) for different array parameters $L_1$-$L_2$ (in nm).

**Fig. 2** Magneto-resistance for different arrays ($L_1$-$L_2$ in nm). SEM images are shown for each array, with the vortex distribution for various *n*. The scale bar corresponds to 200 nm. The matching field $B_\phi$ is indicated (Tesla), as well as $t = T/T_{C0}$ and $J$ in A/m$^2$.

**Fig. 3 (a)** Critical current vs. applied field for different arrays $L_1$-$L_2$ (in nm, see legend). $\Delta J_C$ is indicated. **(b)** Relative $J_C$ enhancement and **(c)** normalized elastic (squares) and pinning (triangles) energy variation as a function of $L_1$. Open symbols for n=2 and solid ones for n=4.

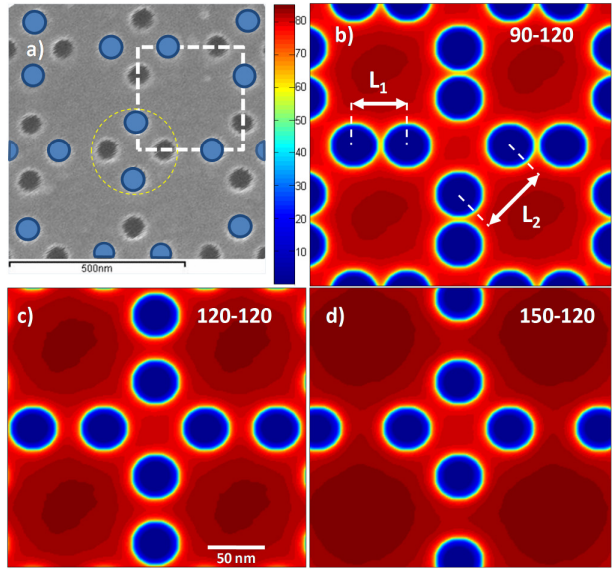

FIG. 1

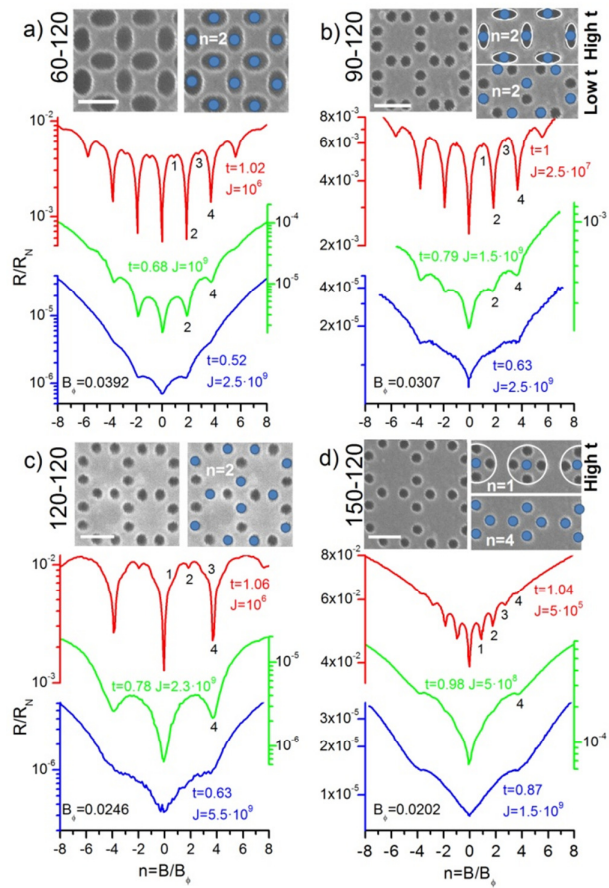

FIG. 2

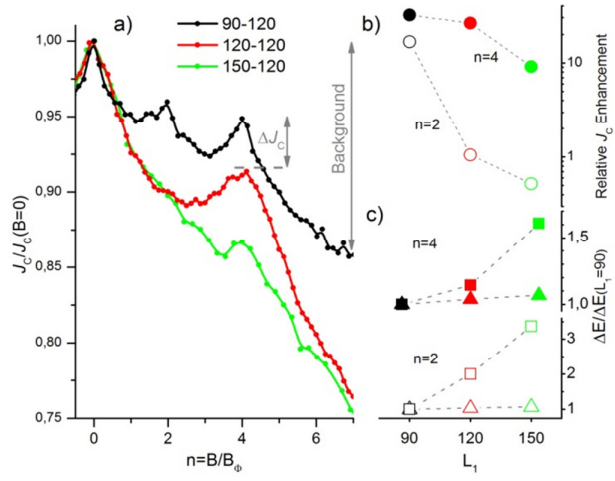

FIG. 3